\documentclass[runningheads]{llncs}
\usepackage{graphicx}

\usepackage{hyperref}
\hypersetup{
    colorlinks,
    linkcolor={red},
    citecolor={black},
    urlcolor={blue}
}

\begin{document}
\title{Nostradamus: Weathering Worth}
%
%
\author{Alapan Chaudhuri\orcidID{0000-0001-4836-9430} \and
Zeeshan Ahmed\orcidID{0000-0002-2719-7355} \and\\
Ashwin Rao\orcidID{0000-0002-1545-2611} \and
Shivansh Subramanian\orcidID{0000-0001-5647-1790} \and\\
Shreyas Pradhan\orcidID{0000-0002-3762-3653} \and
Abhishek Mittal\orcidID{0000-0001-8987-8386}}
\authorrunning{Chaudhuri et al.}
%

\institute{IIIT Hyderabad, Telangana 500032, India
\email{
\{alapan.chaudhuri,zeeshan.ahmed\}@research.iiit.ac.in, 
\{shivansh.s,shreyas.pradhan\}@research.iiit.ac.in, 
\{ashwin.rao,abhishek.m\}@students.iiit.ac.in}
}
\maketitle              
\begin{abstract}
Nostradamus, inspired by the French astrologer and reputed seer, is a detailed study exploring relations between environmental factors and changes in the stock market. In this paper, we analyze associative correlation and causation between environmental elements (including natural disasters, climate and weather conditions) and stock prices, using historical stock market data, historical climate data, and various climate indicators such as carbon dioxide emissions. We have conducted our study based on the US financial market, global climate trends, and daily weather records to demonstrate a significant relationship between climate and stock price fluctuation. Our analysis covers both short-term and long-term rises and dips in company stock performances. Lastly, we take four natural disasters as a case study to observe the effect they have on people’s emotional state and their influence on the stock market.

\end{abstract}
\section{Introduction}
A financial market \cite{1} is an aggregation of buyers and sellers trading stocks or their derivatives. A popular saying in the context of trading stocks goes that \textit{people decide the worth of the company}. Though it aptly describes the definition, we often see that such is not the case. Due to various (expected or unexpected) reasons, companies end up deviating from market trends. Our goal here is to look into such deviations which might be resulted from climate conditions, weather or natural calamities and establish their statistical significance so that we can predict, analyse and appropriately react to such events in the future.

\section{Dataset}

\subsection{Data Extractions}
We have taken historical stock data for multiple companies using Yahoo Finance API \cite{2} and ZEPL US Stock Market Data \cite{3}. We have also extracted sustainability data about each company from Yahoo for general usage and insights. All of the data is publicly available for free and is also stored on our \href{https://github.com/banrovegrie/nostradamus}{GitHub}.
For environmental Data, we have used KNOEMA Environmental Data Atlas \cite{4} to get environmental data for multiple countries over many years. We have got $40+$ parameters related to CO$_2$ emissions, fuel consumption, and Use of nitrogen, amongst others per country per year. To get climate-related data, we have used NOAA Climate Data \cite{5}, which gives us the temperature, precipitation and snowfall of a given area per day. 

 \subsection{Data Cleaning and Enhancements}
Once we have obtained the data required as data-frames, we performed feature selection to reduce the noise in the data. This made our process more efficient and decreased space limitations. We have performed two different analyses: yearly and daily. For yearly data analysis, we have grouped the weather data, climate data and historical OHLC data by year to aptly represent the columns. For daily data analysis, we computed Moving Averages, Volume Weighted Average Prices \cite{6}, Uncertainty and 50 Day Standard Deviation. 

\begin{equation}
    VWAP = \frac{\sum_0^{n-1} Volume_i \times \frac{High + Low + Close }{3}}{\sum_0^{n-1} Volume }
\end{equation}

\begin{equation}
    SD = \sqrt{\frac{\sum_0^{n-1} \mid x - \bar{x} \mid}{n}}
\end{equation}

We have stored all the data required in separate CSV files to decrease the runtime. We also created a pipeline which enables the user to fetch all the data for any stock they require easily. 
 
\section{Environmental Factors}

We hypothesised that there are multiple environmental factors which affect the government and the people directly, but not the companies so much. Companies change their policies, which change their revenues, based on reactions from the government and their customers. In a graph, this would be reflected by an initial change in trend (based on people's opinion), which later settles a bit and then changes based on whether or not government changes their policies. 

\begin{equation}
    r_{xy} = \frac{\sum (x_i - \bar{x}) (y_i - \bar{y})}{\sqrt{\sum (x_i - \bar{x})^2 \sum (y_i - \sum\bar{y})^2}}
\end{equation}

Recent work in this direction includes studies by Castro et al. 2021 \cite{7} and Zahid et al. 2022 \cite{8}. They looked majorly into the effects on the European financial markets and South Asian financial markets respectively.

\subsection{Method}
We define the value of correlation \begin{math} r_{xy} \end{math} as the ratio of the sum of deviations divided by the root of the product of the sum of squares of deviations. A high value of the correlation index (close to 1) indicates that the environmental factor and the company's profits share a causal relationship, whereas a correlation index closer to -1 represents an inverse of such a relationship.

\begin{figure}[htbp]
\centerline{\includegraphics[scale=0.125]{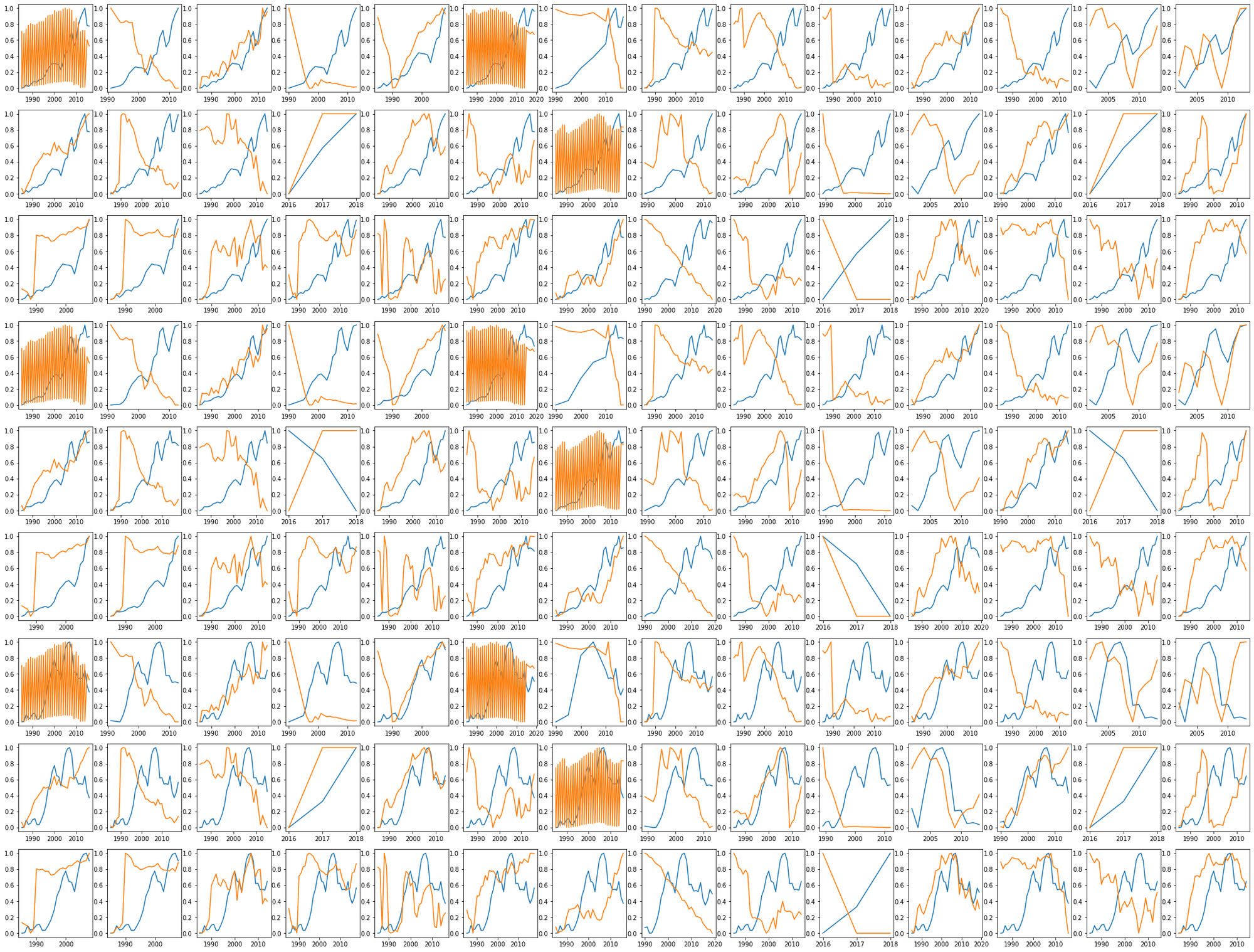}}
\label{fig}
\caption{Represents correlation between various companies' stock prices vs CO$_2$ emission in the USA over the years (view full-sized image in our \href{https://github.com/banrovegrie/nostradamus}{GitHub repository}).}
\end{figure}

\subsection{Inferences}

Multiple stocks show a different correlation between CO$_2$ emissions and stock prices, and each of them signifies important information related to their sector.

\subsubsection{Strong Correlations} A high positive value of correlation usually means that there is an interdependence (Strong Correlation) between a stock and an environmental factor. Consider the stock  (British Petroleum Company), which is an oil and gas company. The company's production has a direct effect on CO$_2$ emissions. Hence, we can confidently infer that CO$_2$ emission values are affected by BP's stock. 

\begin{figure}[htbp]
\centerline{\includegraphics[scale=0.5]{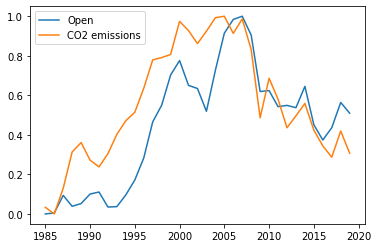}}
\label{fig}
\caption{Shows the correlation between open prices of BP over the years vs CO$_2$ emissions in the USA. Correlation value as high as 0.8159 was observed.}
\end{figure}

\subsubsection{Causation versus Correlation} It is not always true that a high value of correlation means that there is an interdependence between stock prices and CO$_2$ emissions. Consider the stock of AAPL (Apple) and the CO$_2$ emissions from gaseous fuel consumption. The correlation value observed was close to 0.93. We know that Apple (a tech company) is not dependent on CO$_2$ emissions from  gaseous fuel consumption. Despite that, it has a high correlation with that factor. This is not a result of dependence between the two things. The high correlation is simply a coincidence as CO$_2$ emissions from gaseous fuels are rising because of the rapid population growth and because nuclear and other clean energy sources are not very prevalent. Hence \textit{correlation} is not always a result of \textit{causation} \cite{9}.

\begin{figure}[htbp]
\centerline{\includegraphics[scale=0.5]{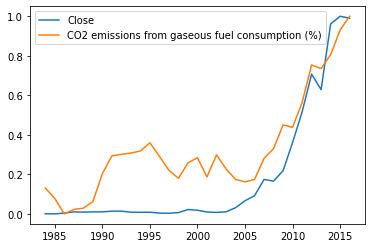}}
\label{fig}
\caption{Shows the correlation between open prices of AAPL over the years vs CO$_2$ from gaseous fuel consumption in the USA. Correlation value as high as 0.93 was observed.}
\end{figure}

\subsubsection{Hidden Correlations} Sometimes a stock and an unrelated environmental factor have a high correlation. We observed that there are $2$ possible explanations for it: either it is just a coincidence (as shown above), or it has a hidden correlation. For example, consider the stock of EOD (Wells Fargo Global) vs CO$_2$ emissions in the USA. The graph has a high positive correlation value of 0.927. However, there are likely hidden correlations due to dependencies of carbon emissions on an industry which also determines the prices of companies which own/invest in the same. This includes large companies such as large banks and firms. Hence, this high correlation is very likely not a result of coincidence but a result of purposeful investing/decisions taken by the company.

\begin{figure}[htbp]
\centerline{\includegraphics[scale=0.5]{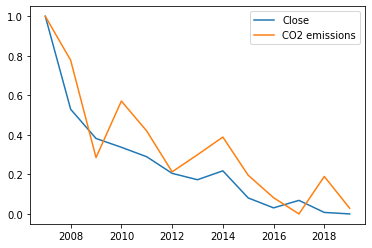}}
\label{fig}
\caption{Shows the correlation between open prices of EOD over the years vs CO$_2$ emissions in the USA. Correlation value as high as 0.927 was observed.}
\end{figure}

\subsubsection{Low Positive Correlations} A low positive correlation usually means that the stock and the environmental factor in consideration are independent of each other. Consider the case of XOM (Exxon Mobil) and Agricultural Methane Emissions, which has a low correlation of 0.234. Hence, we can most of the time conclude that these two are independent of each other. In rare cases, though, they could be dependent and still have a low correlation.

\begin{figure}[htbp]
\centerline{\includegraphics[scale=0.5]{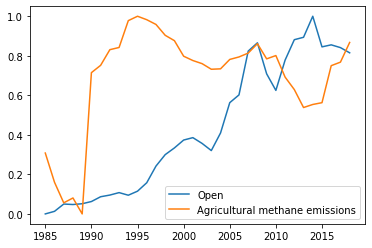}}
\label{fig}
\caption{Shows the correlation between open prices of XOM over the years vs Agricultural Methane Emissions in the USA. Low correlation value of 0.234 was observed.}
\end{figure}

\subsubsection{High Negative Correlations} Highly negative correlation values generally imply an inverse effect between the company's production/success and the factor in consideration. Consider the volume of the stock CVX (Chevron Corporation) and the factor of Terrestrial and marine protected areas. This has a highly negative correlation of -0.899. It is fair to assume that when the number of terrestrial and marine protected areas increases, the volume of the stock CVX, an energy industry, decreases. Hence, a highly negative correlation value means inverse dependence.

\begin{figure}[htbp]
\centerline{\includegraphics[scale=0.5]{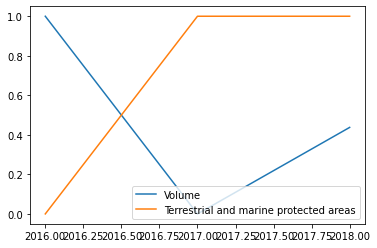}}
\label{fig}
\caption{Shows the correlation between open prices of CVX over the years vs Terrestrial and Marine Protected areas in the USA. High negative correlation value of -0.899 was observed.}
\end{figure}

\subsection{Predictions}

\begin{figure}[htbp]
\centerline{\includegraphics[scale=0.4]{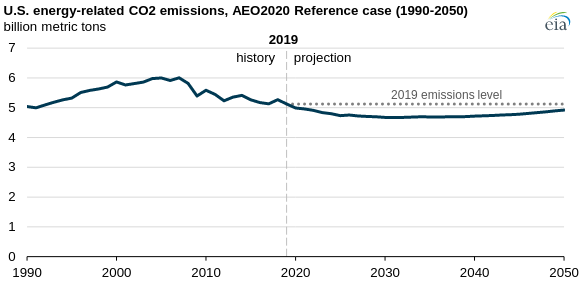}}
\label{fig}
\caption{US energy-related CO$_2$ emissions}
\end{figure}
\noindent The global emission levels are predicted to be stranded due to growing awareness about climate change and a noticeable switch to renewable sources as primary energy sources.

As we earlier saw the high correlation between the carbon emission levels and the stock price of BP, it can be analyzed that in the coming times, the stock price of BP and other large oil companies will begin to fall. This also means that there will be a rise in the stock prices of companies that provide an alternative source of fuels, such as solar panels and windmills.

\subsection{Industrial Revolution}
The example of correlation and not causation can further be understood by considering the rise in the population and workforce. The rise in population meant more consumption of energy, and since we lack any large source of energy other than fossil fuels, it means that there will be a rise in pollution levels.



\begin{figure}[htbp]
\centerline{\includegraphics[width=\textwidth]{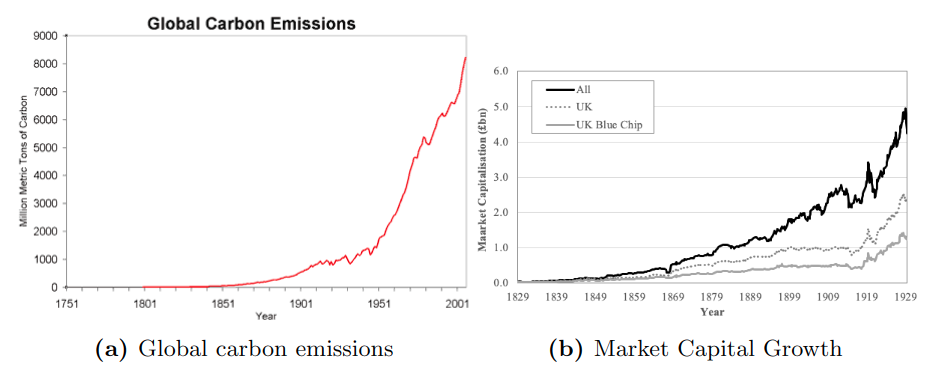}}
\label{fig}
\caption{Correlations between markets and carbon emissions.}
\end{figure}





If the stock history of the industrial revolution times and the carbon emission levels of the same were available, it could be seen that the rising workforce and demand directly meant a rise in carbon levels. This is a correlation due to familiar sources of the rise.

\section{Natural Disasters}

Now, that we have covered aspects of environmental factors, we wanted to experiment with the effects of sudden calamities on the stock market. Therefore, we performed case studies on four disasters (both natural and man-made) occurring in the USA and how prices of certain relevant stocks reacted to them. There have been extensive previous studies in this direction as well. Our conclusions in this respect do concur with several of such works including Tavor \& Teitler-Regev 2019 \cite{10}, Nguyen \& Chaiechi 2021 (who specifically looked into the Hong Kong Stock Market) \cite{11} and Seetharam 2017 (a relatively older paper with an extensive statistical report) \cite{12}.

\subsection{California Wildfires}

In terms of property damage, 2017 was the most destructive wildfire season on record in California then, surpassed by only the 2018 season, with 9,560 fires burning \cite{13}. Throughout 2017, the fires destroyed or damaged more than 10,000 structures in the state (destroyed 9,470, damaged 810).


\begin{figure}[htbp]
\centerline{\includegraphics[width=\textwidth]{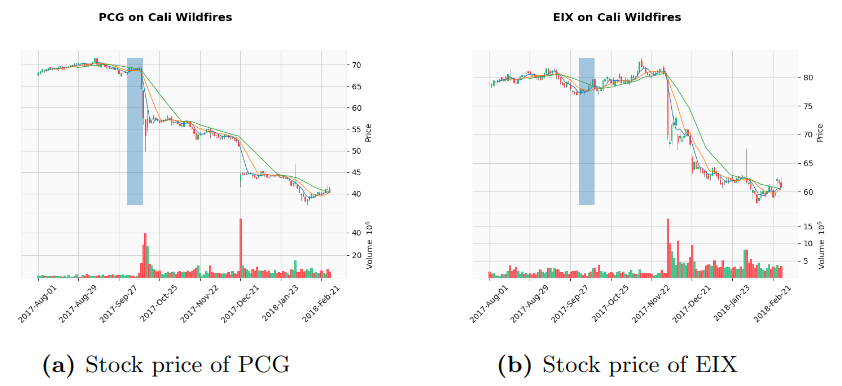}}
\label{fig}
\caption{Stock prices during the 2017-18 California wildfires. The blue bar represents when the wildfires started.}
\end{figure}



We can observe an interdependence between the graphs in this case. The stock of PCG, which is a gas and electric company, was affected right after the disaster took place. But now consider EIX. Its stock fell sharply sometime after PCG's. We know that EIX depends on PCG for its production. This is also evident through the graph as its stock was not affected directly after the wildfires but a few days later. This gives clear information about the dependence of companies on each other.

\subsection{Texas Storm}
In February 2021, the state of Texas suffered a major power crisis \cite{14}, which came about as a result of three severe winter storms sweeping across the United States. The storms caused a massive electricity generation failure in the entire state leading to shortages of water, food, and heat. More than 4.5 million homes and businesses were left without power, and at least 210 people were killed directly or indirectly, with some estimates going as high as 702. The Texas Grid failure was majorly caused by the inadequately winterized natural gas equipment. NRG suffered the most, as can be seen in the following plots. It can also be emphasized that the analysis has to be done on a small time scale as companies often bounce back from such losses over a more extended period of time, such as a year.

\begin{figure}[htbp]
\centerline{\includegraphics[scale=0.15]{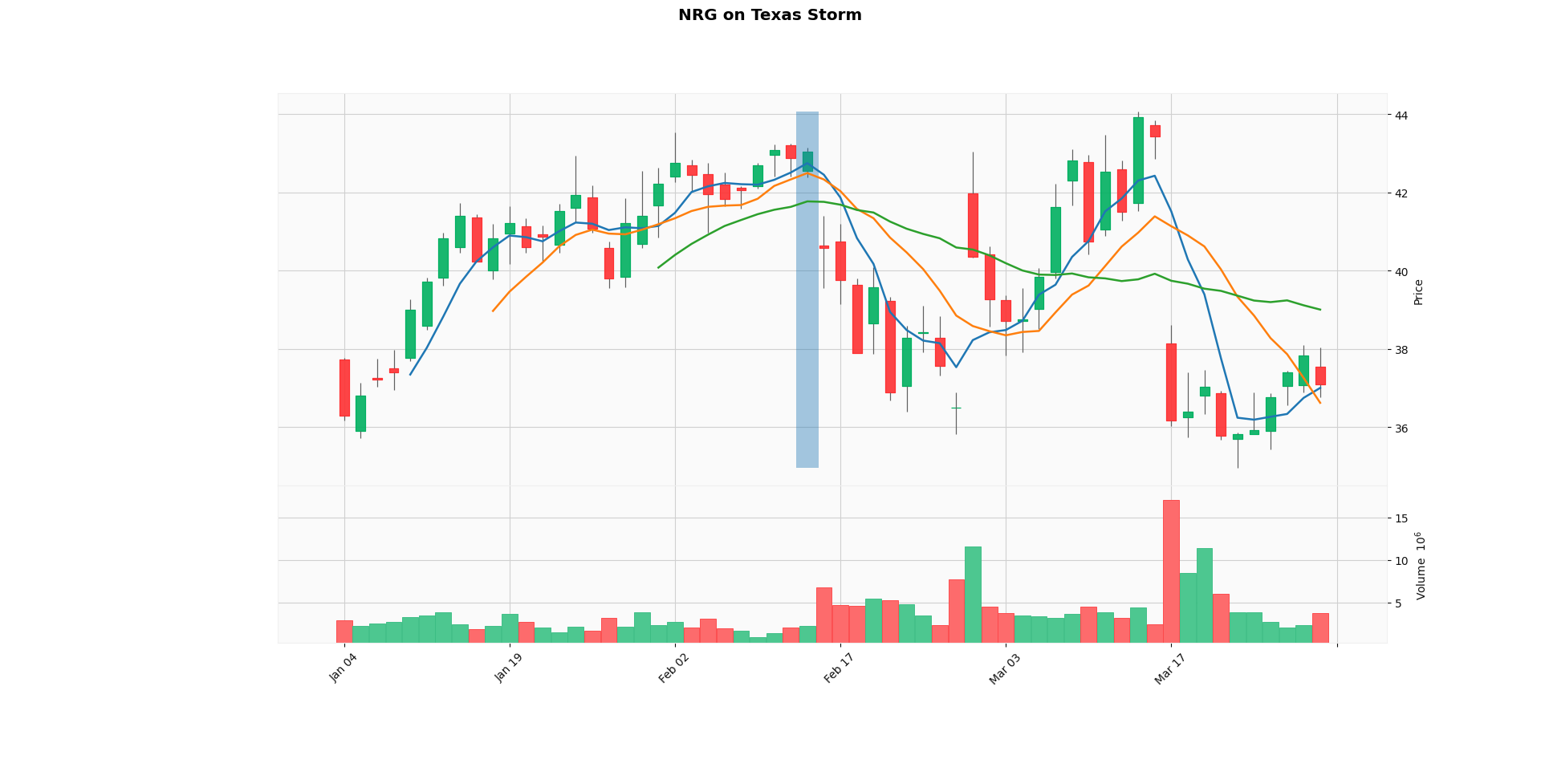}}
\label{fig}
\caption{Stock prices of NRG during the 2021 Texas winter storm. The blue bar represents when the winter storm was at its peak.}
\end{figure}

\subsection{Katrina Hurricane}

In August 2005, Katrina \cite{15} was a category 5 Atlantic hurricane that caused over 1,800 deaths and \$125 billion in damage. This damage was mainly focused on New Orleans and the surrounding areas. Large transportation companies such as C.H. Robinson faced heavy losses due to the denial of services. Since hurricanes have a relatively minor effect on companies, they can resume their services. so the loss is made back.


\begin{figure}[htbp]
\centerline{\includegraphics[width=\textwidth]{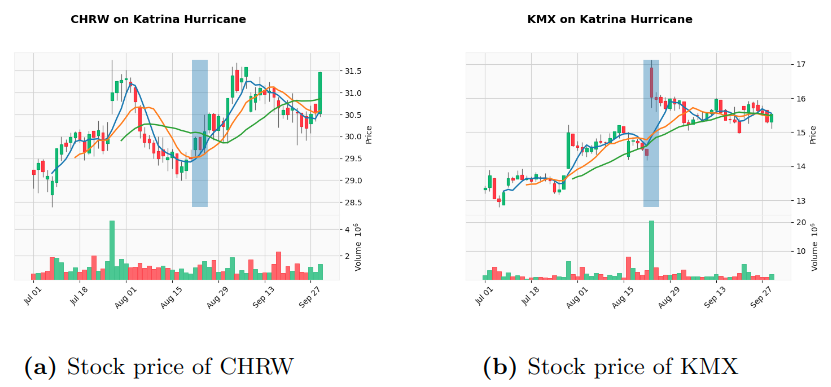}}
\label{fig}
\caption{Stock prices during the 2005 Katrina Hurricane. The blue bar represents when the hurricane occurred.}
\end{figure}



As is seen here, companies related to travel and transport profited here due to the urgent requirement for vehicles for relief and rescue. 

\subsection{9/11}

Although 9/11 \cite{16} was not a natural disaster, it is worth considering how it affected the stock market. The effect is visible on the entire stock market due to direct or indirect loss. As seen in the bottom graphs, the prices go down relatively fast after the event. Due to its surprising nature, the impact's nature and magnitude were no less than a natural disaster over the country.

For all these disasters, stocks of companies related to those events were considered. We observe that AMK, a company related to airlines, suffered huge losses. Even stock indices suffered gap losses representing the general mood of the market and traders after the disasters when everything seemed unsure. In agreement with our hypothesis, the stocks of those companies suffered a heavy hit after those events.

\begin{figure}[htbp]
\centerline{\includegraphics[width=\textwidth]{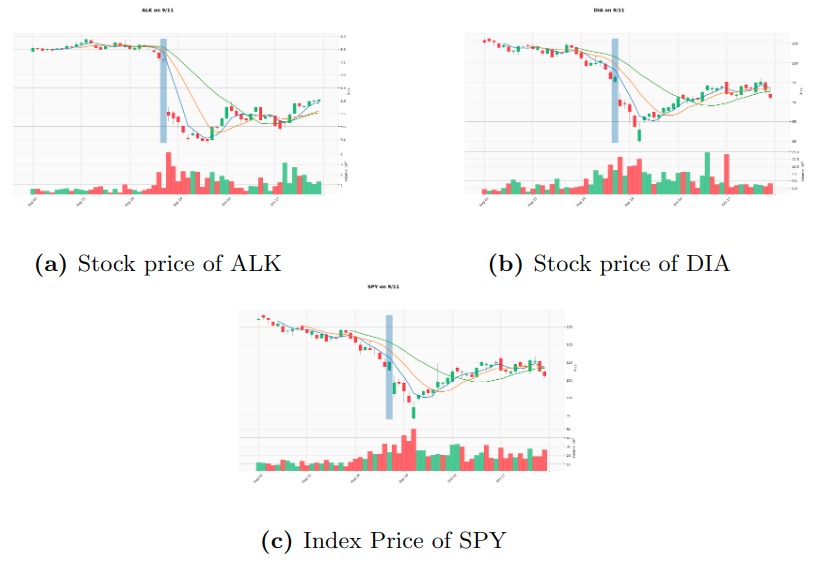}}
\label{fig}
\caption{Stock prices during 9/11. The blue bar represents when the event occurred.}
\end{figure}

\section{Weather}

We try to observe a correlation between daily weather and overall daily trades and obtain a value of $0.0338$, which can be interpreted as statistically insignificant. The stock data we have used for this analysis is SPY, and the weather tracked is of New York City. Our result affirms that of Pardo \& Valor, 2003 \cite{17} and Wang \& Kuang-Hsunshih \& Jang, 2017 \cite{18}. We also found no correlation between trading volume and precipitation, which supports the conclusion of Wang \& C. Lin \& J. Lin, 2011 \cite{19}. However, this result does contradict several much older works such as Saunders, 1993 \cite{20} and Hirshleifer \& Shumway, 2003 \cite{21}. A possible reason behind such disagreement might lie in the fact that we are dealing with market data from a much more recent time period. A lot has changed since the 1990s, and early 2000s especially because electronic trading has made the financial markets much more efficient, transparent and location-independent \cite{22}.

\begin{figure}[htbp]
\centerline{\includegraphics[scale=0.08]{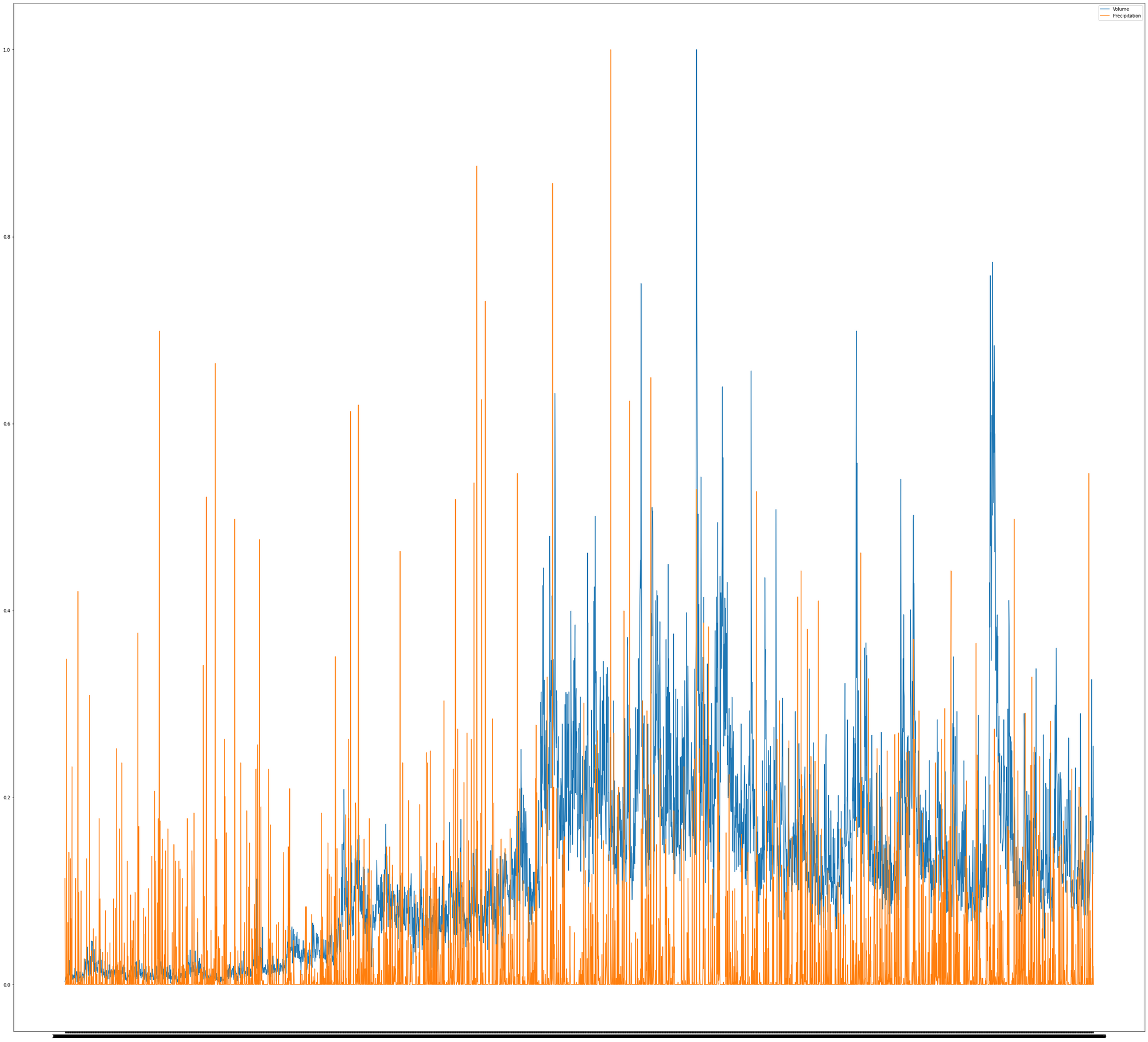}}
\label{fig}
\caption{No particular correlation between weather (NYC daily precipitation and temperature composite in orange) and daily trading (a composite index representing SPY volatility and volume in blue).}
\end{figure}

\section{Conclusion and Future Work}

Our analysis categorically characterises correlations between stock prices and environmental factors. We can conclude with high confidence that even though short-term changes in daily weather generally have no effects on daily trades at the present time, relatively unpredictable events, including natural disasters have a significant toll on stock prices. Interestingly not all stocks get affected negatively.

Furthermore, we have studied the financial and economic ties associated with impactful environmental factors and the cyclicity of catastrophes. Events like hurricanes and heat waves recur periodically. Due to rising climate change, they will only become more frequent. This means that past analysis of such events will help us determine the kind of impact companies would be likely to face. We hope this work would help in being a stepping stone towards more studies in this direction. A promising way forward would lie in building a robust predictive model based on analysis which has already been performed.

%
%
%
%

\end{document}